\documentclass[aps,prl,twocolumn,showpacs,superscriptaddress,groupedaddress]{revtex4} 

\usepackage{graphicx}
\usepackage{dcolumn}
\usepackage{bm}

\usepackage[english,english]{babel}
\usepackage{amsmath}
\usepackage{amssymb,amsfonts,textcomp}
\usepackage{array}
\usepackage{supertabular}
\usepackage{multirow}
\usepackage{placeins}
\usepackage{hhline}
\usepackage[usenames]{color}
\usepackage{soul}
\usepackage{textcomp}
\usepackage{graphicx}
\usepackage{float}
\usepackage{dcolumn}
\usepackage{bm}
\usepackage{comment}
\usepackage{color}
 \usepackage{float}
\usepackage{times}

\begin{document}

\preprint{PRL} 

\title{Using the Marked Power Spectrum to Detect the Signature of Neutrinos in Large-Scale Structure}

\author{Elena Massara}
\email{elena.massara.cosmo@gmail.com}
\affiliation{Waterloo Centre for Astrophysics, University of Waterloo, 200 University Ave W, Waterloo, ON N2L 3G1, Canada}
\affiliation{Center for Computational Astrophysics, Flatiron Institute, 162 5th Avenue, New York, NY 10010, USA}

\author{Francisco Villaescusa-Navarro}
\affiliation{Department of Astrophysical Sciences, Princeton University, Peyton Hall, Princeton NJ 08544, USA}
\affiliation{Center for Computational Astrophysics, Flatiron Institute, 162 5th Avenue, New York, NY 10010, USA}%

\author{Shirley Ho}
\affiliation{Center for Computational Astrophysics, Flatiron Institute, 162 5th Avenue, New York, NY 10010, USA}%
\affiliation{Department of Astrophysical Sciences, Princeton University, Peyton Hall, Princeton NJ 08544, USA}%
\affiliation{Department of Physics, Carnegie Mellon University, Pittsburgh, PA, USA}%
\author{Neal Dalal}
\affiliation{Perimeter Institute for Theoretical Physics, 31 Caroline Street North, Waterloo, ON N2L 2Y5,
Canada}

\author{David N. Spergel}
\affiliation{Center for Computational Astrophysics, Flatiron Institute, 162 5th Avenue, New York, NY 10010, USA}
\affiliation{Department of Astrophysical Sciences, Princeton University, Peyton Hall, Princeton NJ 08544, USA}

\date{\today}

\begin{abstract}
Cosmological neutrinos have their greatest influence in voids: these are the regions with the highest neutrino to dark matter density ratios. The marked power spectrum can be used to emphasize low density regions over high density regions, and therefore is potentially much more sensitive than the power spectrum to the effects of neutrino masses.
Using 22,000 N-body simulations from the Quijote suite, we quantify the information content in the marked power spectrum of the matter field, and show that it outperforms the standard power spectrum by setting constraints improved by a factor larger than 2 on all cosmological parameters.
The combination of marked and standard power spectrum allows to place a $4.3\sigma$ 
constraint on the minimum sum of the neutrino masses with a volume equal to 1 (Gpc $h^{-1}$)$^3$ and without CMB priors. Combinations of different marked power spectra yield a $6\sigma$ constraint within the same conditions.
\end{abstract}

\maketitle


\paragraph{\label{par:intro} Introduction}--- 
Neutrinos are the last particles of the Standard Model whose masses remain unknown. Oscillation experiments have measured two nonzero mass splittings among active neutrinos, showing that at least two mass eigenstates have nonzero mass, but the absolute mass scale and the ordering of the eigenstates remain unknown (see \citep{deSalas:2017kay} for a recent review).  Upcoming laboratory experiments (e.g tritium endpoint and double beta decay experiments) are expected to improve bounds on the neutrino mass scale (\cite{Drexlin:2013lha} for review).

In the near future, cosmology offers a promising independent probe of neutrino masses \citep{Zeldovich1981, Lesgourgues2006, Dvorkin2019}. Neutrinos are so abundant in the universe that their collective mass affects the growth of cosmological structure, producing  distinctive signatures detectable with upcoming surveys.  Cosmological large-scale structure (LSS) is very sensitive to the sum of the masses, which is $M_\nu = \sum_i m_i > 0.06$ eV if two neutrinos are light and one massive (normal hierarchy), or $M_\nu > 0.1$ eV if two neutrinos are massive and one light (inverted hierarchy).  The current tightest constraint comes from combining observations of the cosmic microwave background anisotropies with baryonic acoustic oscillation measurements,  $M_\nu <0.12$ eV at $95\%$ C.L.\ for a flat $\Lambda$CDM cosmology \cite{Planck:2018}. 

If the late time matter/galaxy density fields were Gaussian, then all the cosmological information would be embedded in their two-point functions. Non-linear gravitational evolution generates small-scale non-Gaussianity, inducing an information leakage from the two-point function to higher order statistics (see \cite{Hahn:2019zob,Chudaykin:2019ock,Coulton:2018ebd} for discussions on the bispectrum). 
One way to retrieve this lost information is utilization of different summary statistics, e.g.\ statistics of peaks or voids. Voids have not undergone virialization and are thus expected to retain much of their initial cosmological information~\cite{Pisani:2019cvo}. 
Voids are especially appealing as probes of neutrino physics: since they are much emptier in baryons and dark matter than they are in neutrinos, voids are the regions where the ratio between the cosmic neutrino density and the cold dark matter density is the highest in the Universe~\cite{Massara:2015msa}. Recently, \cite{Massara:2015msa,Banerjee:2016zaa,Kreisch:2018var}  discuss the sensitivity of void-related observables to neutrino masses; \cite{Paco2020} uses the Quijote simulations to estimate the information content of upcoming LSS surveys.

The power spectrum
is the most commonly used observable to extract cosmological information from large-scale structure. 
Since the density power spectrum is significantly affected by the most massive objects \cite{Rimes2005}, it is expected to be sub-optimal when extracting information embedded in low density regions such as cosmic voids. Here we consider a way to use power spectra 
that gives more weight to low density regions, by utilizing 
the so-called marked power spectrum \cite{Stoyan1984}.
For the first time, we explore using marked statistics to weigh neutrinos.

\begin{figure}
\includegraphics[width=1\columnwidth]{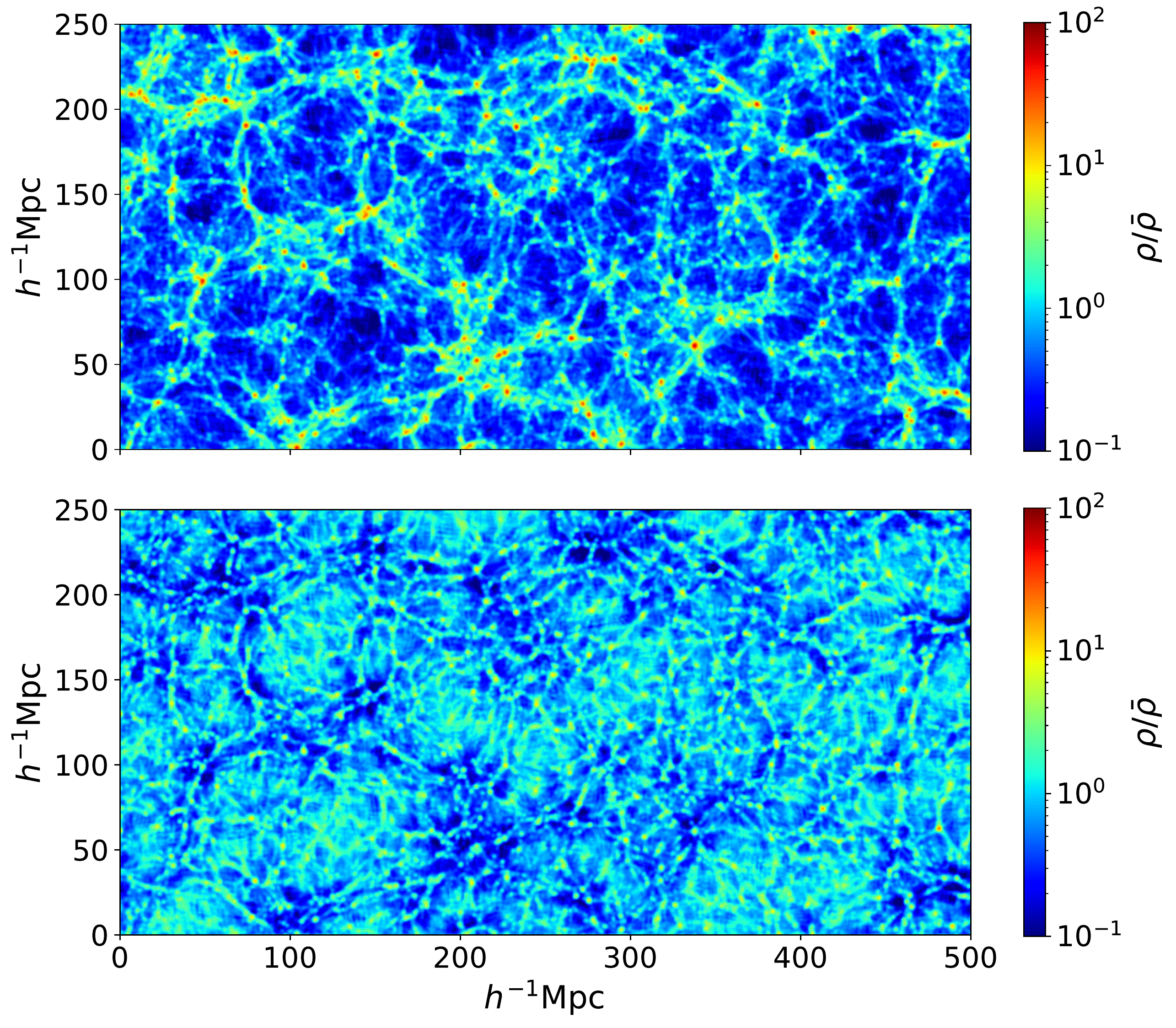}
\caption{\label{fig:density} Projections in a region of $500\times 250\times 20 (h^{-1}{\rm Mpc})^3$ of a simulation with fiducial cosmology at $z=0$. Top panel shows the projected matter density field and bottom panel displays the marked density field with parameters $R=10~h^{-1}$Mpc, $p=2$, and $\delta_s = 0.25$. Here we show how the large-scale structure looks like in a marked field with positive $p$: high density regions are down-weighted, while low density regions are up-weighted.}
\end{figure}

\paragraph{\label{par:Mk} Marked power spectrum}--- Marked correlation functions are modified two-point correlation functions where pairs are weighted by a mark.
The mark usually depends on properties of the considered tracer or on environment. Correlations of marked point processes have been firstly formalized by \cite{Stoyan1984}, and they have been subsequently used in astrophysics to study how the galaxy clustering depends on galaxy properties such as morphology, luminosity, color, etc.\ \cite{Beisbart_2000,Sheth_2005,Skibba_2005}, and how halo clustering depends on merger history \cite{Gottloeber_2002}. A description of marked correlation function in the framework of the halo model has been developed in \cite{Sheth_2005_hm}. More recently, \cite{White_2016} proposed a mark that depends on local density with the purpose of studying modified gravity models. This mark aims to increase the weight of pairs in low density regions, where modifications of gravity are more likely to be present, and it has been used by \cite{Valogiannis_2017,Armijo_2018,Hernandez-Aguayo_2018}. 
Here we consider the mark proposed in \cite{White_2016},
\begin{equation}
    m(\vec{x};R,p,\delta_s) = \left[ \frac{1+\delta_s}{1+\delta_s+\delta_R(\vec{x})}\right]^p\, ,
    \label{eq:mark}
\end{equation}
to build a marked statistics of the matter density field. The mark depends on the local density $\delta_R(\vec{x})=\rho_R(\vec{x})/\bar{\rho}-1$, where $\bar{\rho}$ is the mean matter density and $\rho_R(\vec{x})$ is the local density around the position $\vec{x}$ computed by smoothing the matter density field with a Top-Hat filter of radius $R$. Overall, the mark is a function of three parameters: a scale $R$, a density parameter $\delta_s$ and an exponent $p$. The case $\delta_s\rightarrow 0$ is particularly instructive since it yields $m(\vec{x})\rightarrow \left[\bar{\rho}/\rho_R(\vec{x})\right]^p$. In this case it is clear that positive values of $p$ enhance the weight of points in low density regions, while negative values of $p$ weight more points in high density regions. Figure~\ref{fig:density} displays the marked density field in the case of positive $p$ and shows how high/low density regions become down/up-weighted by the mark.  

In configuration space, the marked correlation function is
\begin{equation}
   1+ M(r) = \sum_{i,j=1}^N \frac{\delta_D(|\vec{x}_i-\vec{x}_j|-r)\,m(\vec{x}_i)\,m(\vec{x}_j)}{\bar{m}^2 N^2/V}\, ,
   \label{eq:m_r}
\end{equation}
where $\delta_D$ is the Dirac delta and $\bar{m}$ is the mean value of the mark. 
In the literature, marked correlations are defined as $(1+M)/(1+\xi)$, where $\xi$ is the correlation function. Here we are not interested in removing the spatial clustering of points, and therefore we do not use this definition. 
Moreover, we choose to work in Fourier space rather than configuration space, due to the lower computational cost needed to measure power spectra.

To sum up, the marked power spectra are promising summary statistics to study neutrinos for multiple reasons. They are straightforward to compute: the measurement of the local density field is the only ingredient that needs to be added to a power spectrum pipeline in order to compute marked power spectra. 
Moreover, the marked power spectra contain information from higher order statistics because of the dependence of the mark on the local density. 
One way to see this is to note that when a mark in Eqn.\ \eqref{eq:mark} including the density field is used, then the marked power spectrum is equivalent to the power spectrum of a nonlinear transformation of the density field.  As many previous works have noted, certain nonlinear transformations can make the density field more Gaussian and thereby transfer information from high-order correlations back to the 2-point function, significantly improving parameter 
constraints from 2-point statistics (e.g.\  \citep{Neyrinck2009,Neyrinck2011a,Neyrinck2011b}). 
Finally, marked power spectra overcome the need of identifying voids, which can be computationally costly depending on the void finder used. Yet, they extract the information from low density regions. 

\begin{figure}
\includegraphics[width=0.95\columnwidth]{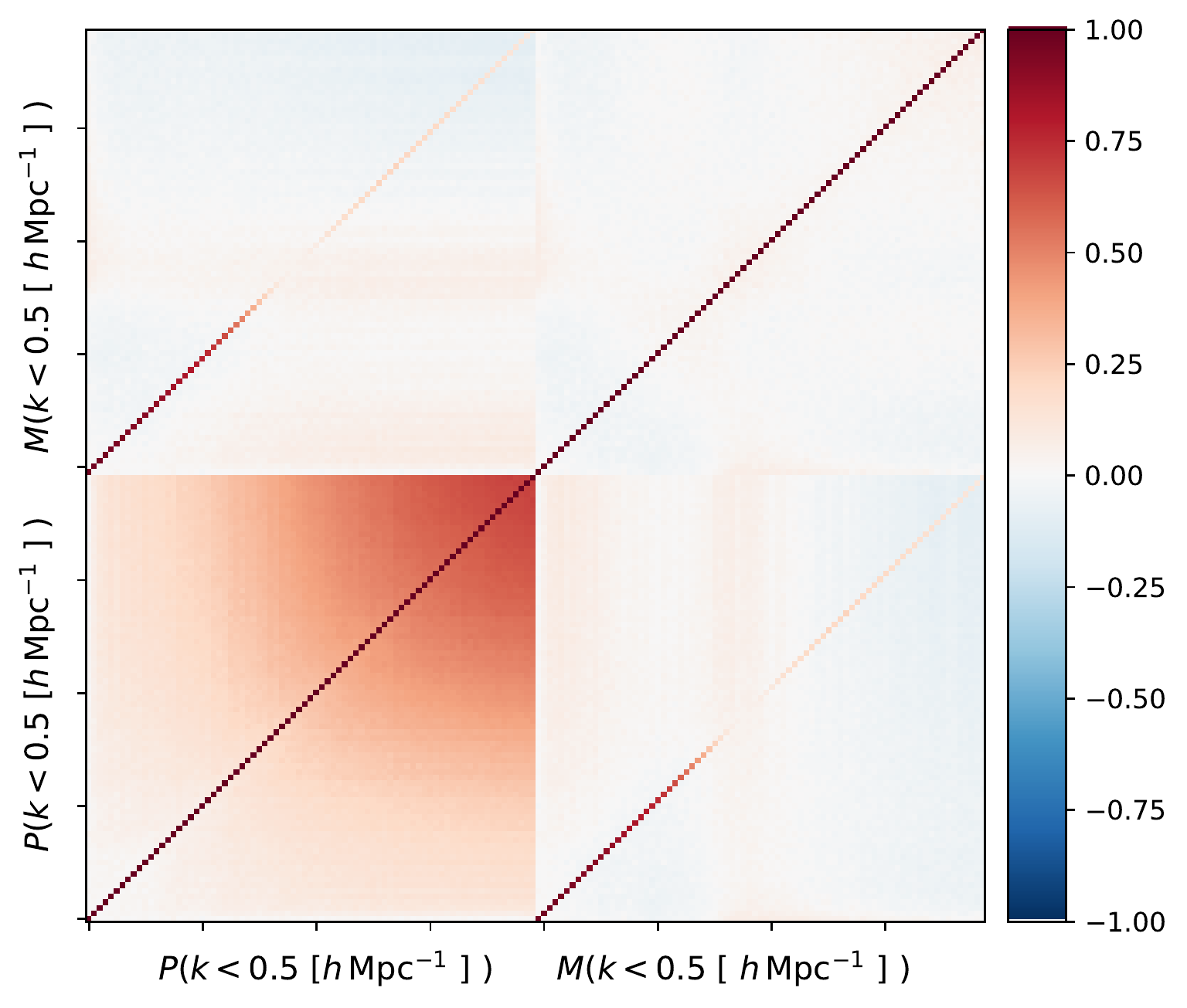}
\caption{\label{fig:corr}Correlation matrix of the power spectrum and of the marked power spectrum with parameters $R=10~h^{-1}$Mpc, $p=2$, and $\delta_s = 0.25$, at $z=0$. The covariance of the latter is clearly more diagonal than the former, which allows for more information to be extracted from small scales. }
\end{figure}

\paragraph{\label{par:fisher} Fisher formalism}--- We quantify the information content (or constraining power) of both standard and marked power spectra using the Fisher formalism. In this framework, the marginalised error squared $\sigma^2(\theta_i)$ associated with a cosmological parameter $\theta_i$ is $\sigma^2(\theta_i) \geq (F^{-1})_{ii}$,
where $F$ is the Fisher matrix defined as
\begin{equation}
    F_{ij} = \frac{\partial \vec{d}}{\partial \theta_i}C^{-1}\frac{\partial \vec{d}^T}{\partial \theta_j} ,
    \label{eq:fisher}
\end{equation}
with $\vec{d}=(O_1,O_2,...)$ being the data vector containing the considered observable $O$ (in our case the marked power spectrum and/or the standard power spectrum) evaluated at different wavelength $k$, and $C$ being the covariance matrix. A large number of simulations are needed to accurately evaluate both the derivatives and the covariance matrix.

\paragraph{\label{par:sims} Simulations}--- The Quijote suite~\cite{Villaescusa-Navarro:2019bje} is a set of more than 43,000 full N-body simulations with different values for the cosmological parameters. Their fiducial cosmology is $\Omega_{\rm m}=0.3175$, $\Omega_{\rm b}=0.049$, $h=0.6711$, $n_s=0.9624$, $\sigma_8=0.834$, $M_\nu=0.0$ eV, and $w=-1$. We used the 15,000 realisations of the Quijote suite run in the fiducial cosmology to compute covariance matrices, and the simulations with one parameter varied above or below its fiducial value to compute numerical derivatives (notice that this analysis focuses on 6 cosmological parameters ($\Omega_m$, $\Omega_b$, $h$, $n_s$, $\sigma_8$ and $M_\nu$) and $w$ is fixed to $-1$). All used simulations contain $512^3$ cold dark matter particles (plus $512^3$ neutrino particles in the simulations with massive neutrinos) in a 1 ($h^{-1}$Gpc)$^3$ box. The derivatives w.r.t.\ $M_\nu$ have been computed using simulations with Zel'dovich initial conditions (see \cite{Villaescusa-Navarro:2019bje} for more details). For each realisation at redshift $z=0$, we measure the matter power spectrum and a set of 125 different marked power spectra. The latter are obtained by considering different values for the three mark parameters:  $R=\left[5,10,15,20,30\right]\, h^{-1}$Mpc, $p=\left[-1,0.5,1,2,3\right]$ and $\delta_s = \left[0,0.25,0.5,0.75,1\right]$. When considering massive neutrino cosmologies, we compute each statistics for both the total matter density field `$m$' (cold dark matter + baryons + neutrinos) and the cold dark matter + baryons density field `$cb$'. We verified that the number of realisations used to compute covariance matrices (15,000) and derivatives (500) gives a convergent estimation of the Fisher matrix and consequently of the errors associated to the cosmological parameters.

\begin{table*}
\begin{ruledtabular}
\begin{tabular}{ccccccccc}
Parameter &	$P_{cb}$ & $M_{cb}$ & $P_{cb}+M_{cb}$ & $M_{cb}+M'_{cb}$ & $P_{m}$  & $M_{m}$  & $P_{m}+M_{m}$ & $M_{m}+M'_{m}$\\
\hline
$\Omega_m$ & 0.046 & 0.018 & 0.017 & 0.014 & 0.094  & 0.013  & 0.012 & 0.011 \\
$\Omega_b$ & 0.016 & 0.0099 & 0.0091 & 0.008 & 0.039  & 0.010  & 0.009 & 0.008\\
$h$ & 0.16 & 0.092 & 0.083 & 0.068 & 0.50 & 0.098  & 0.082 & 0.069 \\
$n_s$ & 0.10 & 0.045 & 0.04 & 0.029 & 0.48  & 0.048  & 0.039 & 0.028\\
$\sigma_8$ & 0.080 & 0.030 & 0.026 & 0.021 & 0.013  & 0.0019  & 0.0015 & 0.0015\\
$M_\nu$ & 1.4 & 0.50 & 0.44 & 0.35 & 0.83 & 0.017  & 0.014 & 0.01 \\
\end{tabular}
\caption{\label{tab:error}Marginalised errors on the cosmological parameters obtained with the Fisher analysis for the standard ($P$) and marked ($M$ and $M'$) power spectrum, and different combinations of them, including the modes with $k<k_{\rm max}=0.5~h{\rm Mpc}^{-1}$. The subscripts $cb$ and $m$ stand for cold dark matter + baryons and matter, respectively.
The inclusion of the marked power spectrum in the Fisher analysis improves significantly the constraints on all cosmological parameters. In particular, the combination of two marked power spectra improves the power spectrum constraints on total neutrino mass $M_\nu$ by a factor of $4$ and $80$ when `$cb$' or `$m$' are considered.
}
\end{ruledtabular}
\end{table*}

\begin{figure*}
\includegraphics[width=2\columnwidth]{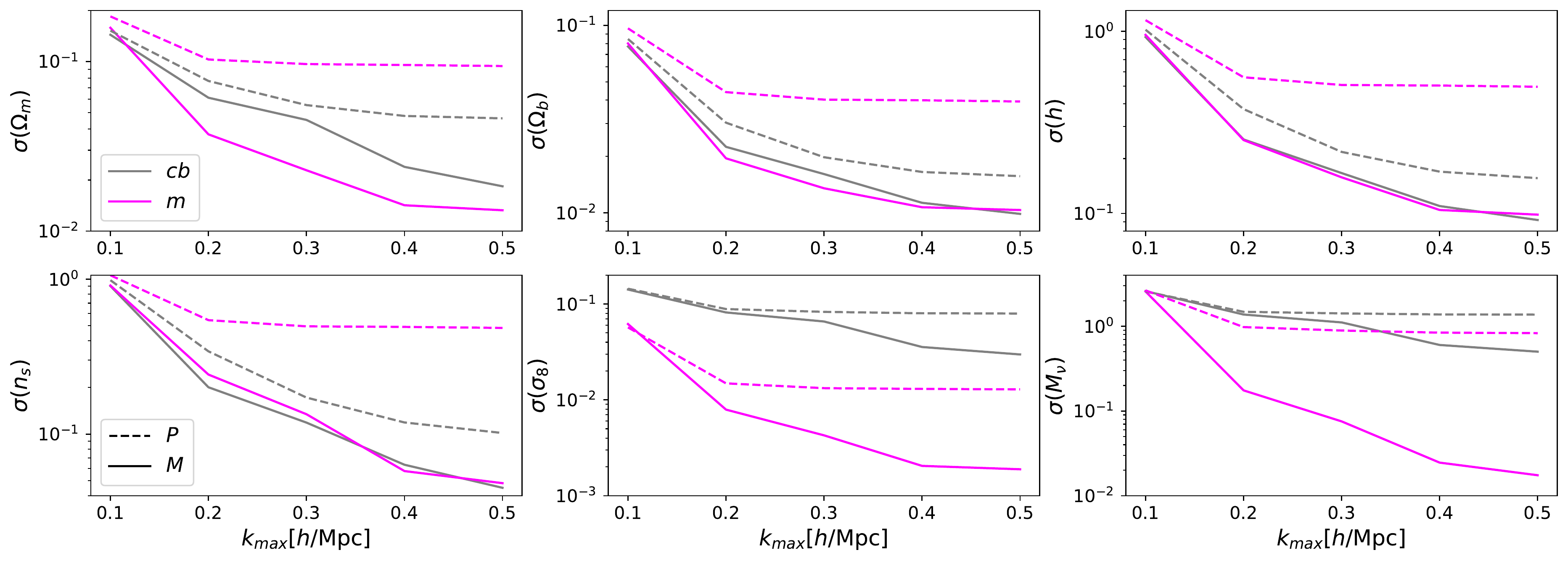}
\caption{\label{fig:errors} Marginalized errors on the cosmological parameters and their dependence on the maximum wavelength considered. Solid lines show the results for the standard power spectrum and dashed lines show the ones for the marked power spectrum $M$ with $R=10~h^{-1}$Mpc, $p=2$, and $\delta_s = 0.25$. The color code refers to the field considered: grey for cold dark matter + baryons and magenta for total matter. The marked power spectra ($M$, solid line) always outperform the power spectrum ($P$). The effect is even stronger when including neutrinos ($M_m$, solid magenta line, compared to $M_{cb}$, solid grey line).}
\end{figure*}

\paragraph{\label{par:results} Results}--- We use the Fisher formalism to quantify the information content on the nonlinear matter power spectrum, and use it as a benchmark to compare to the marked power spectrum. 
We distinguish the case where the considered statistics is computed on the `$m$' or the `$cb$' density field.
The marginalized errors obtained by including all modes with $k<0.5~h$ Mpc$^{-1}$ are presented in Table~\ref{tab:error}. The constraints on the sum of the neutrino masses from the standard power spectrum are $\sigma_{cb}(M_\nu)=1.4$ eV and $\sigma_{m}(M_\nu)=0.8$ eV, where the subscript indicates the field used. As expected, tighter constraints on $M_\nu$ can be achieved when using the total matter power spectrum, because neutrino effects are larger in the `$m$' field than in the `$cb$' field \cite{Massara:2015msa,Banerjee:2019omr}. The marginalised errors as a function of the maximum wavelength included in the analysis are displayed in Figure~\ref{fig:errors}.

The Fisher formalism can be used to identify the mark model --- among the 125 considered models --- that gives the tightest constraint on the neutrino masses. 
In order to avoid the regime where the Quijote simulations may be affected by numerical resolution, we restrict our analysis to the regime where $R\geqslant10~h^{-1}$Mpc. We find that the best value of the mark parameters for both the `$cb$' and `$m$' cases is:
$R=10~h^{-1}$Mpc, $p=2$, and $\delta_s = 0.25$. 
The marginalised errors for this model $M$ are shown in Table~\ref{tab:error}, where all the results have been obtained by considering only modes with 
$k<0.5~h$Mpc$^{-1}$. Even if chosen to maximize the information on neutrinos, the mark power spectrum improves the power spectrum constraints on all the other cosmological parameters by a factor of $2-3$ and $4-10$ when `$cb$' or `$m$' are considered. The errors on the neutrino masses from the marked power spectrum are $\sigma_{cb}(M_\nu)=0.5$ eV and $\sigma_{m}(M_\nu)=0.017$ eV. The latter indicates that marked power spectra on the total matter density can put a $3.5\sigma$ constraint on the minimum sum of the neutrino masses using a volume equal to 1 (Gpc $h^{-1}$)$^3$. Moreover, the chosen marked power spectrum improves the power spectrum constraints on $M_\nu$ by factors of $2.8$ and $47$ when the `$cb$' and `$m$' field are considered, respectively.

This large improvement arises for multiple reasons. First, the covariance matrix of the marked power spectrum is much more diagonal than the one of the power spectrum (Figure~\ref{fig:corr}): this allows extraction of more information on small scales. Figure~\ref{fig:errors} shows that the information in the power spectrum saturates on scales $k<0.5~h$Mpc$^{-1}$, but it does not in the mark power spectrum. Second, the marked power spectrum contains higher order statistics of the density field: \cite{Hahn:2019zob} showed that the bispectrum can improve the power spectrum constraints on $M_\nu$ by a factor $\sim 5$, and \cite{Uhlemann:2019gni} showed more modest improvements using the matter PDF, which also contains higher order statistics. Third, \cite{Banerjee:2019omr} showed that neutrinos induce a large and unique scale-dependent bias on linear scales, when halos/galaxies are split according to neutrino environment. The mark power spectrum could be using these unique features to boost its constraining power. Finally, as stated above, our marked power spectrum has been designed to incorporate information from voids into the power spectrum.

The combination of standard and marked power spectrum measurements allows to obtain tighter constraints on cosmological parameters, and in particular on $M_\nu$: $\sigma_{cb}(M_\nu)=0.44$ eV and $\sigma_{m}(M_\nu)=0.014$ eV ($4.3\sigma$ constraint on the minimum sum of the neutrino masses). Even tighter constraints can be achieved by combining two or more different mark models. For example, the combination of the mark considered above ($M$) and a second one ($M'$) with parameters $R=10~h^{-1}$Mpc, $p=1$, and $\delta_s = 0$ yields $\sigma_{cb}(M_\nu)=0.35$ eV and $\sigma_{m}(M_\nu)=0.01$ eV ($6\sigma$ constraint on the minimum sum of the neutrino masses). The usage of 3 mark models can improve these constraints by a factor of $1.3$ and $1.1$ when considering the `$cb$' and `$m$' fields, respectively.

\paragraph{\label{par:discussion} Discussion}--- 
In this letter we propose the usage of marked power spectra as efficient probes to weigh neutrinos and in general to tightly constrain the value of the cosmological parameters. For the first time, we computed the information content on marked power spectra and compared it with the one from the standard power spectrum. 
We also showed that combinations of different marked power spectra can yield very tight constraints, and these combinations outperform the standard power spectrum in constraining all considered cosmological parameters.  

The analysis has been done at the underlying density field level, distinguishing 
the `$cb$' and `$m$' density fields. The latter brings tighter constraints on the neutrino masses, but it cannot be observed in galaxy redshift surveys directly \citep{Castorina:2013,Villaescusa-Navarro:2013}. One possible way to measure it is through weak-lensing observations, which give a 2D projection of the underlying `$m$' field. The `$cb$' density field cannot be observed directly either, but galaxies and other objects are tracers of it \citep{Castorina:2013,Villaescusa-Navarro:2013}. Combinations of galaxy clustering and weak-lensing measurements can be used to define marks sensitive to `$m$' and `$cb$'.

We have not considered the contribution of super sample covariance to the covariance matrix, which will degrade our constraints. Moreover, when using galaxy clustering, theoretical uncertainties such as galaxy bias, redshift-space distortions, and baryonic effects are also expected to degrade our constraints after marginalizing over them. Our results show that a $6\sigma$ constraint on the minimum sum of the neutrino masses can be achieved by considering combinations of marked power spectra of the total matter density in a volume equal to $1~(h^{-1}{\rm Gpc})^3$. Upcoming surveys such as DESI \cite{DESI}, Euclid \cite{Euclid} and WFIRST \cite{WFIRST} are expected to probe volumes of tens of $(h^{-1}{\rm Gpc})^3$. Thus, these surveys should achieve a statistically significant detection of the neutrino masses, even if a significant fraction of the information content is lost when marginalizing over theory uncertainties \footnote{We defer for an upcoming work a throughout analysis of the impact of theory uncertainties on the results of this work.}. We emphasize that these constraints will arise solely from large-scale structure surveys, without the usage of CMB priors. Thus, they will complement the results of CMB constraints \citep{Simons2019,CMBS4} and serve as an internal cross-check to verify the robustness of the results.

EM and FVN would like to thank Benjamin D. Wandelt and Martin White for useful discussions. This work has made use of the Pylians libraries, publicly available at https://github.com/franciscovillaescusa/Pylians, and results were obtained using the Gordon cluster in the San Diego Supercomputer Center. This work was partially supported by NASA grant 15-WFIRST15-0008 and NASA ROSES grant 12-EUCLID12-0004. During the realisation of this project EM, FVN, SH and DS were supported by the Simons Foundation.  ND is supported by the Centre for the Universe at Perimeter Institute. Research at Perimeter Institute is supported in part by the Government of Canada through the Department of Innovation, Science and Economic Development Canada and by the Province of Ontario through the Ministry of Economic Development, Job Creation and Trade.


\newcommand{\apjl}{Astrophys.\ J.\ Lett.}
\newcommand{\mnras}{Mon.\ Not.\ Roy.\ Astron.\ Soc.}

\bibliography{Mk_biblio}

\end{document}